# Novel heterojunction bipolar transistor architectures for the practical implementation of high-efficiency three-terminal solar cells

Pablo G. Linares*, Elisa Antolín, and Antonio Martí

Instituto de Energía Solar, Universidad Politécnica de Madrid
Avenida Complutense, 30
28040, Madrid, Spain

**Abstract**

Practical device architectures are proposed here for the implementation of three-terminal heterojunction bipolar transistor solar cells (3T-HBTSCs). These photovoltaic devices, which have a potential efficiency similar to that of multijunction cells, exhibit reduced spectral sensitivity compared with monolithically and series-connected tandem solar cells. In addition, the simplified n-p-n (or p-n-p) structure does not require the use of tunnel junctions. In this framework, four architectures are proposed and discussed in this paper: 1) one in which the top cell is based on silicon and the bottom cell is based on a heterojunction between silicon and III-V nanomaterials; 2) one in which the top cell is made of amorphous silicon and the bottom cell is made of an amorphous silicon-silicon heterojunction; 3) one based on the use of III-V semiconductors aimed at space applications; and 4) one in which the top cell is based on a perovskite material and the bottom cell is made of a perovskite-silicon heterostructure.

*Corresponding author: p.garcia-linares@upm.es

1. Introduction

The future of photovoltaics (PVs) requires maximizing the efficiency, of which multijunction solar cells (MJSCs) currently hold the record [1], and the implementation of low-cost materials and fabrication processes similar to those of the current crystalline Si (c-Si) PV industry. A three-terminal heterojunction bipolar transistor solar cell (3T-HBTSC) has been recently proposed [2] as a PV concept with the potential to combine the high efficiency of III-V MJSCs, which makes efficient use of the solar spectrum [3], and the low-cost of Si or thin-film PVs in a simple and straightforward semiconductor configuration.

In the following, we describe the fundamentals of 3T-HBTSCs, focusing on the main features that make them resemble bipolar junction transistors (BJTs) in terms of semiconductor structure but not device operation. Concerning PV potential, they are compared to MJSCs, describing commonalities in their optical performance and addressing the advantages of the prior in regard to enhancement of the energy production and reduction of the complexity of the semiconductor layer. A number of different 3T-HBTSC architectures based on different materials, configurations and fabrication methods are presented in this work. The new types of solar cells discussed here are also applicable to a number of different PV applications and the corresponding PV markets.

2. Fundamentals of 3T-HBTSC

The 3T-HBTSC design hybridizes optical aspects from MJSCs and electrical features from three-terminal devices in a monolithic configuration that resembles an n-p-n (p-n-p) BJT but does not perform as such. The three-terminal PV device concept is not new [4], although previous approaches were based on the mechanical stacking of single-junction solar cells [5] or on two independent p-n junctions separated by a highly doped layer [6]. The fundamental structure of this solar cell is exclusively based on the three aforementioned n-p-n (or equivalently, p-n-p) semiconductor layers. Fig. 1 (a) shows a schematic representation of this simplified 3T-HBTSC design, in which conventional parts of the solar cell, such as the window, back surface field (BSF) and anti-reflecting coating (ARC) layers meant to improve surface passivation, reduce minority carrier recombination and minimize reflection losses, have not been considered because their contribution does not affect the discussion.

Nomenclature from BJTs is employed in Fig. 1, which shows, from top to bottom, an n-type high-bandgap emitter in orange, a p-type high-bandgap base in blue and a low-bandgap collector again in orange. Two conductive interdigitated grids act as top and middle contacts for the emitter and base layers, which should be properly designed to maximize the sunlight reaching the semiconductor while reducing resistive losses during photocurrent collection. A bottom contact is located at the rear of the collector. The higher energy portion of the solar spectrum is absorbed by the top subcell, which is composed of the top high-bandgap (emitter) and middle (base) high-bandgap n-p junction. Lower energy photons are transmitted to the bottom subcell, constituted by the high-bandgap (base) and low-bandgap (collector) p-n heterojunction, where they are absorbed (by the collector). Fig. 1 (b) shows a sketch of the 3T-HBTSC under illumination, performing similarly to a conventional dual-junction MJSC in the sense that sunlight photons are selectively absorbed by consecutive PV materials with decreasing junction bandgaps. The main differences are the use of only three semiconductor layers and the extraction of the currents generated by each junction through independent circuits instead of through a series-connected configuration. The three-terminal cell is depicted in the figure with a common contact (B) at the base, from which two different external loads can independently bias each subcell. Both the top and bottom subcell photocurrents are extracted through the base terminal (B in the figure), with the first one re-entering the device at the emitter (E) electrode and the second one at the collector (C) electrode. Notice that the output voltage of the top cell is only limited by the high bandgap of the top emitter and base semiconductors, while the output voltage of the bottom cell is limited by the low bandgap of the collector.

Both the top and bottom subcells share the middle layer, which offers a path for the quasi-Fermi level (QFL) split in the bottom cell (i.e., the voltage of the bottom cell) to propagate towards the top cell, as represented in the energy band diagram sketched in Fig. 2 (a), and therefore can potentially limit the open-circuit voltage ($V_{OC}$) of the top cell. The electron and hole QFL are referred to as $E_{Fe}$ and $E_{Fh}$, respectively, in the figure. In previous three-terminal solar cells designed and fabricated to date, this QFL continuity is broken by the insertion of a highly conductive layer (also called a lateral conduction layer or LCL) between the two p-n junctions, so that the structure of the solar cell becomes np-p$^+$-pn [7]. However, according to the new PV paradigm presented in Ref. [2], this p$^+$ layer is not required, provided that the transistor effect in the n-p-n structure is prevented. On this basis, such transistor effect (i.e., the current amplification gain) can be prevented if the middle layer has a higher doping level than the top layer and/or the bandgap of the middle layer is larger than that of the top layer. In this case, the QFL separation at each subcell is independent from the others, as shown in the 3T-HBTSC energy band diagram represented in Fig. 2 (b). The understanding of this effect allows an important simplification of the solar cell structure, as the PV conversion efficiency limit can be dramatically increased, reaching that of a MJSC, by simply adding a single semiconductor layer to a conventional p-n junction.

3T-HBTSCs are n-p-n devices like BJTs; however, to perform correctly, the emitter in the latter must have a higher doping level or a higher bandgap than the base, in contrast to the emitter in the former. Two typical magnitudes describe the operation of a BJT: the transport factor ($\alpha_T$) or ratio between the collector and emitter electron (for an n-p-n transistor) currents ($I_{C,n}$ and $I_{E,n}$, respectively), as shown in Equation 1, and the emitter current or emitter injection efficiency ($\gamma_E$), defined as the electron emitter current divided by the sum of the electron ($I_{E,n}$) and hole ($I_{E,p}$) currents across the base-emitter junction (Eq. 2).

$$\alpha_T = I_{C,n} / I_{E,n} \qquad (1)$$

$$\gamma_E = \frac{I_{E,n}}{I_{E,n} + I_{E,p}} \qquad (2)$$

Ideal BJTs require both $\alpha_T \rightarrow 1$ and $\gamma_E \rightarrow 1$, but 3T-HBTSCs require $\gamma_E$ to be as close to 0 as possible, thus giving a device performance far from that of a "good" transistor [2].

Regarding the implementation of more complex PV devices with increasing efficiency, 3T-HBTSCs can be used as building blocks for more complex structures, for example, by incorporating a higher number of semiconductor materials, so that the overall performance is equivalent to a MJSC with a higher number of junctions.

## 3. Comparison with MJSCs

3T-HBTSCs are compared to MJSCs because they perform similarly regarding the absorption of photons at consecutive subcells made from materials with decreasing bandgaps to which unabsorbed lower energy photons are transmitted. However, conventional MJSCs face the disadvantage of high sensitivity to the spectral illumination conditions imposed by subcell series-connection. This causes the photogenerated current to be wasted in the form of recombination when one or more junctions produce excess current, i.e., under any illumination spectrum other than the one providing current-matching. Moreover, the epitaxial structure is highly complex because of the need for so-called tunnel junctions, which allow series-connection of the different device subcells with minimized voltage loss. The transistor-like devices presented here are meant to perform as double-junction MJSCs without the need for tunnel junctions, thus only employing three semiconductor layers (n-p-n or p-n-p, like BJTs) instead of six layers, corresponding to the two p-n junctions plus the $n^+$-$p^+$ tunnel junction of a conventional two-terminal MJSC. The three-terminal configuration provides an independent circuit for the extraction of each subcell $I_L$, thus preventing the photogenerated carriers from being wasted due to current-mismatched subcells under spectrally unbalanced illumination. The latter affects the static picture not only of a particular illumination but also of any other set of environmental conditions. Therefore, the energy output delivered by a 3T-HBTSC throughout the day is greatly improved under this independently connected scheme. Moreover, throughout the year, this effect is magnified, since the spectral conditions are highly variable due to atmospheric filtering resulting from the inclination of the Sun with respect to a particular location on the Earth's surface [8]. Similar arguments apply to the reduction of the electric production of a PV cell subject to aging: encapsulant material/optics yellowing and other age-driven alteration of the nominal device features that predominantly affect one of the junctions do not impact the whole 3T-HBTSC $I_L$, in contrast to conventional MJSCs. In this regard, the higher robustness against environmental degradation shown by the three-terminal cells can be particularly relevant in space applications, where high-energy particle radiation, known to severely damage only one of the MJSC subcells [9] does not reduce the photogenerated current produced by the non-degraded subcell in a 3T-HBTSC. Other optical mechanisms, such as luminescent coupling [10] which also depends on the electrical bias point, can easily remove the conventional MJSC from its current-matched regime, thus preventing photon recycling. Conversely, 3T-HBTSCs exploit this effect efficiently, since the extra photogenerated current added by the lower bandgap subcell can always be extracted through their independently connected terminals.

Among MJSCs, four-terminal (4T) cells have several similarities with our 3T-HBTSC: the two subcell photocurrents are also extracted independently and there is no need for a tunnel junction. However, the cell processing of the 4T is more complicated since: a) it demands two interdigitated metal grids instead of one in order to implement the fourth contact which, in addition, have to be perfectly aligned to prevent extra shadowing of the bottom junction and b) it requires a minimum of four semiconductor layers instead of three. On the other hand, both in the 3T as in the in the 4T approaches the current in the middle layers (the base in the case of the 3T) would have to be extracted laterally which represents, in this respect, a common design problem for both architectures.

Not all of the distinct features of 3T-HBTSCs can be regarded as an advantage over those of MJSCs. Efficient three-terminal cell connection into modules is challenging. Recently it has been demonstrated a two-terminal module prototype comprising HBTSCs which provides a high-voltage power output [11]. That module works optimally in the case voltage ratio between the top subcell and the bottom subcell equal two. For an arbitrary voltage ratio the module interconnection remains an engineering challenge to be solved by voltage-conversion electronics. Furthermore, the integration of 3T-HBTSCs with increasing bandgaps, forming an equivalent quadruple- or sextuple-junction cell with a higher limiting efficiency, has yet to be demonstrated.

4. The c-Si approach

Two approaches have been identified regarding the possibility of manufacturing a 3T-HBTSC based on c-Si. The first approach requires a low-bandgap heterojunction to be formed at the bottom of a standard c-Si solar cell to extend the absorption of infrared photons from the solar spectrum. As an example, an In(Ga)As nanowire (NW)-c-Si cell is proposed as a candidate architecture of this type.

The second approach relies on the addition of a higher bandgap junction on top of a standard c-Si cell. As an example, a Si heterojunction with intrinsic thin-layer (HIT) cell could be modified so that an extra hydrogenated amorphous Si (a-Si:H) layer is included on top, thus absorbing part of the high-energy solar photons at a significantly higher voltage. Both structures, together with their potential advantages as well as their implementation strategies, will be reviewed in the next sections. The details of the processing steps required to fabricate the two aforementioned 3T-HBTSC structures based on c-Si are presented in the appendix at the end of the article.

4.1. c-Si/In(Ga)As NW cells

State-of-the-art Si solar cells, with a record efficiency of 26.3% [12], based on an interdigitated back contact comprising an a-Si:H heterojunction are highly optimized, and their performance is close to the theoretical efficiency limit (29.1%). Meanwhile, the growth of free-standing InAs NW on Si has been verified [13] experimentally, showing phase-pure zinc blende crystal structures with the expected very low bandgap slightly above 440 meV and the potential to create a high-quality Si/In(Ga)As NW p-n junction with few defective regions. A scanning electron microscope (SEM) image of the InAs nanostructures with NW widths of only tens of nanometers grown on c-Si by molecular beam epitaxy (MBE) is shown in Fig. 3 (a). Other studies [14] have shown the potential of c-Si/InAs NW junctions grown by metal−organic chemical vapor deposition (MOCVD) for PV applications. Furthermore, the addition of Ga for the fabrication of InGaAs NW structures on c-Si layers has also been demonstrated [15], enabling controllable tuning of the bottom heterojunction subcell bandgap and the corresponding modification of its absorption range. State-of-the-art Si solar cell technology can be inherited by the three-terminal cell architecture proposed here, so that the addition of a low-bandgap III-V nanomaterial in the form of an extra In(Ga)As NW junction at the bottom transforms the n-p c-Si junction into an n-p-n 3T-HBTSC, thus improving the efficiency without compromising the performance of the highly efficient Si PV cell. The use of III-V-based NWs allows the formation of a crystalline p-n junction on c-Si with a large lattice mismatch [13]. In(Ga)As NWs were chosen because of their very low bandgap, starting from approximately 0.4 eV, i.e., a few tens of meV above that of bulk InAs.

The simplified layer architecture of an example n-p-n Si/In(Ga)As NW 3T-HBTSC is shown in Fig. 3 (b), where each color represents a different semiconductor material corresponding to the three junctions and the external contacts: orange for the Si base/substrate, red for the Si (rear) emitter, blue for the In(Ga)As NWs and yellow for each of the three corresponding metal contacts. As indicated in the figure, the rear n-In(Ga)As NW layer is interdigitated, i.e., it does not cover the entire back face of the structure, so that the intermediate p-type Si layer can be externally contacted.

A cost analysis of the Si/In(Ga)As NW solar cell manufacturing process, considering the supplementary expenses of III-V nanomaterial deposition and corresponding middle layer contact processing, shows that upgrading an industrial state-of-the-art c-Si solar cell into a Si/In(Ga)As NW 3T-HBTSC requires at least two additional efficiency points to make it cost effective due to the increased cost with respect to the standard c-Si cell. Although pure InAs NWs could be used for this purpose, the addition of Ga further boosts the performance of the bottom Si/InGaAs heterojunction subcell. Our calculations estimate that the range of InGaAs NW bandgaps capable of such efficiency increase varies from 0.45 to 0.7 eV when the Si/InGaAs NW subcell is illuminated with an AM1.5G solar spectrum and filtered by the higher bandgap c-Si top subcell. This corresponds to an InGaAs ternary cell with a Ga content between 10% and 40%, where the optimum composition can yield a photogenerated current of up to 15 mA/cm$^2$. These calculations are also subject to the c-Si top subcell thickness, which can also be tuned. This novel solar cell architecture has the potential to impact the PV market, which constantly seeks further cost reductions [16] and where most PV modules are still based on c-Si, by means of a considerable efficiency boost at a moderate extra cost.

Fig. 3 (c) outlines the manufacturing steps identified for the addition of the In(Ga)As NW junction. Starting with an n-type Si wafer (1), a p-n junction is formed (2). Following the process proposed in reference [13], a thin SiO$_x$ layer is sputtered on the p-side (3). The wafer is etched in HF. Favored by the inhomogeneity of the sputtered layer, holes are opened on the SiO$_x$ layer, exposing the p-Si underneath (4). At this point, the wafer is introduced in the MBE reactor, and crystalline In(Ga)As NWs are grown on the openings (5). Finally, conventional photolithography is used to define the photoresist patterns used to selectively etch SiO$_x$, which allows access to the intermediate p-Si layer and subsequent deposition of the metal contacts for both the rear p-Si and the n-In(Ga)As NWs, as well as for the metal grid corresponding to the front n-Si (6).

4.2. a-Si:H/Si HIT

Another straightforward 3T-HBTSC architecture based on a c-Si wafer can be implemented by upgrading the HIT Si solar cell through the addition of an a-Si:H top layer. The a-Si:H/Si HIT 3T-HBTSC consists of an a-Si:H homojunction top subcell, with a typical bandgap of 1.75 eV, and a very efficient HIT Si bottom subcell, where most of the absorption takes place within the thick c-Si wafer, i.e., above 1.12 eV. This structure, sketched in Fig. 4, has a vertically aligned double front grid contact, corresponding to both the upper n-type and the intermediate p-type a-Si:H. In this case, the third contact is made on the rear side, covering the entire c-Si layer.

The current know-how of a-Si:H and Si HIT technologies, not only on the laboratory scale but also at the industrial scale, assures high efficiency and low manufacturing cost. High-quality commercial HIT cells could be used as the starting point for this type of 3T-HBTSC to provide a proof of concept, since only minor modifications are required. The high absorption coefficient of the direct-bandgap a-Si:H eases the implementation of this novel structure, because very few micrometers of material are necessary for the absorption of a significant part of the incoming sunlight in this direct-bandgap semiconductor. Compared with a state-of-the-art Si HIT cell, the performance of the Si HIT bottom subcell, corresponding to the a-Si:H/Si HIT 3T-HBTSC, is degraded because of the absorption of high-energy photons at the less efficient a-Si:H homojunction. However, this loss is compensated by the extra gain added by the higher voltage achieved by the a-Si:H top subcell. The HIT structure has attracted much research attention lately because of the low-temperature processes required for its fabrication and the corresponding high potential for reducing manufacturing costs together with the very high performance demonstrated to date [17]. The simplified three-terminal architecture proposed here to implement a multiple-bandgap solar cell based on a HIT configuration represents a straightforward pathway for boosting the HIT solar cell market.

5. High-efficiency III-V solar cells for CPV and space applications

III-V PVs are a mature technology that have dominated the PV efficiency race for more than thirty years and have demonstrated efficiencies above 40%, by the optimization and testing of metamorphic MJSCs under concentrated light [18], for more than a decade. The inherent drawback of massive III-V MJSC deployment is the high manufacturing cost resulting from the III-V or Ge substrates [19], together with the time-consuming high-technology tools required for the epitaxy and processing of III-V solar cells. In this respect, an important research interest is the possibility of manufacturing high-efficiency III-V MJSC approaches using low-cost Si substrates [21], along with substrate reuse strategies [19]. Regarding current applications of III-V MJSCs, two main markets exist: concentrator photovoltaics (CPV), where small cells receive the high-power density sunlight collected by large surface optical elements such as mirrors or lenses, and space satellite missions, where high-efficiency solar panels fabricated from large-area solar cells are employed to supply power. In the space market, the light weight and reliability of each mission spare part are put ahead of cost due to launch weight constraints and the need to minimize the risk of power supply deficiencies. With respect to terrestrial PVs, the high cost of III-V MJSCs in \$/m$^2$ can be mitigated by employing concentrator systems that provide a high solar irradiance, since only a small fraction of semiconductor material is required compared with the total solar panel surface exposed to the Sun, thus drastically reducing the cost in \$/W. Several three-terminal architectures are proposed here for the implementation of III-V-based transistor-like solar cells suitable for use under the AM0 spectrum as well as under concentrated illumination in a CPV system.

GaAs and Ge wafers are often used in industrial state-of-the-art MJSCs and therefore attract high scientific as well as commercial interest. The reasons are two-fold: on the one hand, III-V cells represent, by far, the most efficient PV technology to date [1], and the simplifications proposed by the three-terminal architectures can lead to significant cost reductions with a great potential on the CPV market [22], and on the other hand, the greater spectral flexibility and robustness against outer-space degradation offered by 3T-HBTSCs can be crucial in space missions.

3T-HBTSCs can be implemented with (Al)GaAs/Ge on a Ge substrate as well as with GaInP/(In)GaAs and AlGaAs/GaAs on a GaAs substrate. All these combinations are suitable candidates with optimized efficiency limits, either under the extraterrestrial AM0 spectrum or under a wide range of sunlight concentrations under the AM1.5D terrestrial solar spectrum. 3T-HBTSCs can also be fabricated using III-V materials with a relatively simple and straightforward design, such as that shown in Fig. 5. The schematic of an epitaxial layer structure of an AlGaAs/GaAs cell is represented with its three main transistor-like regions (collector, base and emitter), two front and one rear metal contacts, and other layers meant to improve the solar cell efficiency: BSF and window layers to reflect minority carriers and reduce recombination, a contact layer to improve the metal-semiconductor resistance, and an ARC to minimize reflection losses. III-V alloy stoichiometries, doping levels and thicknesses are indicated for most layers. Notably, both the n-type doping level and the Al content of the emitter layer are lower than those of the base so that the transistor effect is minimized.

Under this configuration, the series resistance caused by the lateral flow of current can limit the cell performance unless independent interdigitated metal fingers contacting the emitter and base reduce the mean length required for majority carrier collection. In this configuration, the trade-off between front grid finger pitch and shadowing factor becomes critical. This is due not only to the need for a double metallic grid on the front side of the cell but also to the wider portion of the emitter that has to be etched out (for the fabrication of the base contact) as a safety margin to prevent shortcuts between the exposed edge of the etched emitter and the base metal fingers. Although the extra shadowing produced by the safety margins only affects photogenerated current collection at the top subcell, the shadowing factor can be large compared with that produced by the metallic fingers. This is particularly relevant for concentrator cell designs where the finger width has to be minimized so that their number can be maximized, thus reducing their pitch and the corresponding lateral resistance. This extra loss can be minimized by means of a double rear contact, which in turn improves the lateral resistance, since there is no shadowing effect on that side of the cell and the metal covers most of the surface (with the exception of the aforementioned safety margins). However, this implementation complicates the cell processing of direct-bandgap III-V semiconductor structures because the active part of the solar cell is only a few micrometers thick, and therefore, the whole substrate would have to be etched from the rear side.

6. Low-cost perovskite cells in high-efficiency configurations

Perovskites are organic-inorganic compounds with a $MABX_3$ crystalline structure, where MA is usually a large methylammonium cation ($CH_3NH_3^+$), B is a smaller divalent cation, typically $Pb^{2+}$, and X is a halogen anion ($I^-$, $Cl^-$ or $Br^-$) bonded to both A and B. The perovskite crystal most often used for PV applications is methylammonium lead triiodide ($CH_3NH_3PbI_3$), as shown in Fig. 6 (a). This crystal forms a cubic-symmetry structure, where $Pb^{2+}$ has 6-fold coordination, surrounded by $I^-$ in an octahedral configuration of anions, and the MA cation appears in 12-fold cuboctahedral coordination.

Perovskite technology has recently gained prominence in the PV realm [23] as a promising material capable of merging dye-sensitized and organic solar cell technologies into an efficient, relatively simple, and inexpensive PV concept. Perovskite materials rely on organic compounds with a high self-organization potential in the nanoporous $TiO_2$ selective contacts employed in dye-sensitized solar cells. Therefore, perovskites are used in device configurations that inherit some of the main technological features from dye-sensitized cells proposed by Brian O'Regan and Michael Grätzel almost three decades ago [24]. This way, perovskite PV cells rely on photon absorption creating an excited state in the perovskite material and subsequent electron transfer to $TiO_2$ and hole transfer to a hole-transporting medium (HTM). In addition to PV recombination losses, perovskite cells are affected by charge transfer at the interfaces between $TiO_2$ and the perovskite, the HTM and the perovskite, and $TiO_2$ and the HTM.

The first perovskite solar cell was presented in 2006 by Kojima et al., who reported an efficiency of 2.2% with a $CH_3NH_3PbBr_3$ perovskite and an HTM made with a lithium halide organic electrolyte [25]. The electrical performance soon evolved when Br was replaced by I and the electrolyte was replaced with a more efficient and stable solid-state spiro-OMeTAD [26]. Recent progress in perovskite solar cells has led to a sharp increase in the conversion efficiency, which currently surpasses 22% [27]. However, this technology still faces intricate challenges, such as the substitution of Pb by a less toxic element and device stability over time [23]. Despite the promise of single-junction perovskites, there is increasing interest in their hybridization with Si in a monolithic multijunction configuration [28]. Their wide tunable bandgap offers very high limiting efficiencies calculated under the Shockley-Queisser limit [29], reaching 45.3% for a 1.81 eV perovskite/c-Si MJSC with three- or four-terminal configuration under 1 sun illumination. This limiting efficiency remains above 40% for a wide bandgap range, from 1.35 eV to 2.4 eV, as shown in reference [30]. Conversely, the maximum efficiency of a series-connected perovskite/c-Si dual-junction cell drops almost 10 points when the perovskite bandgap is varied by only 200 meV. Nevertheless, the current experimental record efficiency of 23.6% [31] indicates the potential of this hybridized structure.

As a consequence of the potential high efficiency at reduced cost offered by perovskites, many research groups worldwide are currently examining this promising concept. To this respect, the three-terminal approach we propose allows a considerable simplification of the perovskite/c-Si structure, as illustrated schematically in Fig. 6 (b), where a 3T-HBTSC architecture adapted to this concept is shown. This architecture is based on a double-superimposed front grid separately contacting emitter and base, which are contacted through indium tin oxide (ITO). The intermediate contact is patterned on top of a $SiO_2$ insulator deposited on the c-Si wafer. Further details of the fabrication process are given in the appendix.

Regarding the rationale behind three-terminal solar cell architectures, perovskite/c-Si 3T-HBTSCs adapt better to device degradation or/and variation in the illumination spectrum during operation compared with conventional multijunction implementations. In addition, a greater flexibility is provided for the choice of the perovskite material with the appropriate bandgap-matching towards any particular spectrum during the design stage. Notably, this configuration avoids the fabrication of a Si p-n junction, and therefore, its implementation mostly involves low-thermal, low-cost processes.

7. Conclusions

In summary, we presented a novel solar cell architecture based on a three-terminal configuration that resembles a BJT, and we proposed its implementation with different PV technologies. 3T-HBTSCs can contribute to state-of-the-art PV technology by 1) improving the performance of current flat-plate PVs beyond 30% efficiency through simple and cost-effective modifications of the c-Si solar cell architecture; 2) increasing the robustness against outer-space degradation with respect to the III-V MJSC used in space applications and maximizing the energy production because of the greater flexibility against variations in the solar spectrum in terrestrial CPV applications – in both cases, the solar cell performance improvements are achieved by employing a simplified layer architecture with an efficiency equivalent to current world-record double-junction cells; [1] and 3) enabling simple and straightforward MJSC-like configurations for low-cost PV concepts that employ novel organic materials with the potential for high conversion efficiency, such as perovskites, combined with well-known c-Si technology.

The first approach shown in this work is based on conventional c-Si solar cells, where a low-bandgap III-V In(Ga)As NW layer is added on the rear side to absorb low-energy photons that cannot be absorbed by the Si cell, thus enhancing its performance without interfering with the operation of the Si cell. The other Si-based 3T-HBTSC relies on a HIT architecture, where the top a-Si:H emitter is capped with another a-Si:H layer, thus producing a high-bandgap (1.75 eV) junction on top of the HIT cell in a straightforward implementation. 3T-HBTSCs have also been proposed for use under concentrated light and in space applications. In both applications, very efficient III-V materials must be used, with which extremely high efficiencies can be obtained with a simplified architecture that reduces the number of layers from 6 to 3 for a dual-junction solar cell. The PV architectures presented here are particularly suitable for space applications, where high-energy radiation is known to degrade one of the junctions in particular, which has a large impact on the overall device performance of conventional two-terminal architectures but not of three-terminal architectures. Finally, perovskite/c-Si 3T-HBTSCs are proposed as an easy and straightforward way to implement low-cost PV devices with the potential for high efficiency. Detailed descriptions of all device implementations are given in the paper.

Appendices

Further details of the processing methods required for the different 3T-HBTSC architectures are compiled here.

**A. c-Si/In(Ga)As NW cells**

Since HIT solar cells are currently the highest efficiency Si technology, they are good candidates for the high bandgap subcell. As sketched in Fig. 3 (c), a plausible design for this 3T-HBTSC would require a top metal grid on the front surface, contacted by a very thin 10 nm n-type emitter with a $10^{19}$ cm$^{-3}$ doping, followed by a high quality 10 nm thick intrinsic layer and a relatively thin 100 μm n-type Czochralski c-Si substrate with a resistivity of ~ 1 Ωcm$^2$, also as part of the emitter and meant to reduce bulk recombination and provide a good enough lateral resistance. The base is buffered by another 10 nm intrinsic layer and a highly doped ($10^{20}$ cm$^{-3}$) 25 nm boron-doped p-type nanocrystalline Si, which provides the high bandgap of the base which prevents the transistor effect. Heavily n-doped In$_{0.7}$Ga$_{0.3}$As NW arrays with a peak energy of 0.6 eV and an areal density of ~ 1.4 $10^8$ cm$^{-2}$ are used as collector of the 3T cell. The manufacturing steps identified for the addition of the In(Ga)As NW junction include (1) deoxidation of an n-type Si wafer; (2) formation of a p-n junction; (3) sputtering of a thin SiO$_x$ layer on the p-side following the process proposed by Koblmüller et al. [13]; (4) HF etching of the wafer so that holes are opened on the SiO$_x$ layer, favored by the inhomogeneity of the sputtered layer, thus exposing the p-Si underneath; (5) MBE growth of crystalline In(Ga)As NW on the openings; and (6) conventional photolithography to define the photoresist patterns used to selectively etch SiO$_x$ to allow access to the intermediate p-Si layer, followed by deposition of the metal contacts for both the rear p-Si and the n-In(Ga)As NW layers, as well as for the metal grid corresponding to the front n-Si.

**B. a-Si:H/Si HIT cells**

Fig. 4 shows a sequence of plausible manufacturing steps for the implementation of the a-Si:H/Si HIT 3T-HBTSC. Essentially, the new structure only requires an additional n-type a-Si:H layer on top of the HIT cell. Regarding thickness and doping of the layers, the high bandgap junction works as a typical a-Si:H, where the intrinsic layer is now designed to absorb high energy photons. Therefore, the upper n-type a-Si:H layer (namely, emitter, in our design) is extremely thin (15 nm) and highly doped (1 $10^{19}$ cm$^{-3}$); the subsequent i-layer is 250 nm thick, enough to ensure the absorption of most of the light above 1.75 eV; the 25 nm thick nanocrystalline Si base, with a doping concentration close to $10^{20}$ cm$^{-3}$, also produces a noticeably high bandgap. After that, a high-quality 10 nm thick intrinsic amorphous layer has to be followed by the aforementioned 100 μm n-type Czochralski c-Si substrate with a resistivity of ~ 1 $\Omega cm^2$ acting as the collector of our 3T-HBTSC. The main processing steps required for its fabrication are described as follows: (1) first, selected areas in the n-Si wafer are protected with $SiO_2$; (2) an intrinsic (i)-a-Si:H layer is deposited; (3) a transparent conductive oxide (TCO), such as indium tin oxide (ITO), is placed above the areas covered with $SiO_2$; (4) a p-type a-Si:H layer, finalizing the bottom Si HIT subcell, is grown; (5) an n-type a-Si:H layer completes the top subcell a-Si:H p-n junction and another layer of ITO is deposited on top of it; and (6) in the last step, the top n-type a-Si:H layer is contacted with a front grid aligned to the previously deposited ITO intermediate grid, so that it shades the poorly passivated c-Si of the HIT bottom subcell. A rear contact covering the whole n-type c-Si is also metalized.

### C. III-V cells

To fabricate a III-V 3T-HBTSC, once the epitaxial structure is grown, the following main processing steps are required: (1) front etching of the cell until the base is reached, (2) intermediate contact deposition, (3) annealing of the intermediate contact, (4) deposition of the front and rear contacts, and (5) annealing of the front and rear contacts (the same temperature can be applied in this case because both the emitter and collector layers have the same type of doping and can comprise a highly doped contact layer), followed by mesa etching, contact layer removal and ARC deposition. The most challenging part, besides the initial etching, which requires accurate control of the etching rate, is the annealing of the intermediate metal contact, because it is incorporated on a highly-reactive $Al_{0.5}Ga_{0.5}As$ layer instead of the much more common GaAs contact layer usually employed in AlGaAs devices.

### D. Perovskite/c-Si cells

The steps for the fabrication of the perovskite/c-Si 3T-HBTSC start with the preparation of the Si wafer that will act as collector layer. A p-type Czochralski c-Si substrate with a resistivity of ~ 10 $\Omega cm^2$ can be used for this purpose. To improve light trapping, a wafer with a textured back surface and a polished front surface is preferred. To produce a back contact, a 500 nm Al layer is evaporated on the back surface and subsequently annealed at 700C in a $N_2$ atmosphere. The purpose of the annealing is to produce an ohmic contact and also to improve the silicon carrier life time through a gettering process. To produce the base contact, a SiO$_2$ (80 nm) / ITO (400 nm) grid is patterned on top of the c-Si by photolithography and sputtering. A ~ 1 µm thick methylammonium lead trihalide perovskite (e.g. CH$_3$NH$_3$PbI$_3$ ) base layer is formed on top of that surface by co-evaporation. Then, a hole transporting material layer is produced to act as the emitter in the 3T-HBTSC. For this purpose, a spyro-OMeTAD thin layer is deposited by spin-coating and a 100 nm ITO layer is sputtered on top of it. If it the technology is available, the spyro-OMeTAD layer can be replaced by a sputtered NiO$_x$ layer. This way, degradation caused by sputtering of the ITO onto the hole transporting layer could be avoided, as well as possible homogeneity problems in the spin-coating related to surface roughness. Finally, a metallic grid (e.g. Cr/Au) can be added by photolithopgraphy and evaporation to improve the series resistance of the top contact.

Acknowledgements

This work was supported by MADRID-PV (S2013/MAE-2780) funded by the regional government of Madrid, INVENTA-PV (TEC2015-64189-C3-1-R) funded by Ministerio de Economía y Competitividad and, GRECO (787289) funded by the European Union's Horizon 2020 Program and the ConCEPT Project, funded by the Iberdrola Foundation. E.A. acknowledges a Ramón y Cajal Fellowship (RYC-2015-18539) funded by Ministerio de Economía y Competitividad. P.G.L and E.A. are also thankful for a Research Line Reinforcement Grant from the Universidad Politécnica de Madrid.

**References**

1.     M. A. Green, Y. Hishikawa, E. D. Dunlop, D. H. Levi, J. Hohl-Ebinger, A. W. Y. Ho-Baillie, Solar cell efficiency tables (version 51), Prog. Photovolt. Res. Appl. 26 (2018) 3–12.

2.     A. Martí, A. Luque, Three-terminal heterojunction bipolar transistor solar cell for high-efficiency photovoltaic conversion. Nat. Commun. 6 (2015) 6902.

3.     E. D. Jackson, Areas for improvement of the semiconductor solar energy converter. in University of Arizona Press (1955) 122–126.

4. J. M. Gee, A comparison of different module configurations for multi-band-gap solar cells, Sol. Cells 24 (1988) 147–155.

5. L. D. Partain, L. M. Fraas, P. S. Mcleod, J. A. Cape, High efficiency mechanical stack using a GaAsP cell on a transparent GaP wafer, in: 18th IEEE Photovolt. Spec. Conf., Las Vegas, Nevada, USA (1985) 539–545.

6. M. J. Ludowise, L. A. Larue, P. G. Borden, P. E. Gregory, T. Dietze. High-efficiency organometallic vapor phase epitaxy AlGaAs/GaAs monolithic cascade solar cell using metal interconnects, Appl. Phys. Lett. 41 (1982) 550–552.

7. M. A. Steiner, M. W. Wanlass, J. J. Carapella, A. Duda, J. S. Ward, T. E. Moriarty, K. A. Emery, A monolithic three-terminal GaInAsP/GaInAs tandem solar cell, Prog. Photovolt. Res. Appl. 17 (2009) 587–593.

8. J. Villa, A. Martí, Impact of the spectrum in the annual energy production of multijunction solar cells, IEEE J. Photovolt. 7 (2017) 1479–1484.

9. M. Zazoui, M. Mbarki, A. Zin Aldin, J. C. Bourgoin, Analysis of multijunction solar cell degradation in space and irradiation induced recombination centers, J. Appl. Phys. 93 (2003) 5080–5084.

10. Henry, C. H. Limiting efficiencies of ideal single and multiple energy gap terrestrial solar cells, J. Appl. Phys. 51 (1980) 4494–4500.

11. M. Zehender, E. Antolín, P, G. Linares, I. Artacho, I. Ramiro, J. Villa, A. Martí, module interconnection for the three-terminal heterojunction bipolar transistor solar cell, in: 14th Intl. Conf. Conc. Photovolt. Syst., Puertollano, Spain (2018).

12. K. Yoshikawa, H. Kawasaki, W. Yoshida, T. Irie, K. Konishi, K. Nakano, T. Uto, D. Adachi, M. Kanematsu, H. Uzu, K. Yamamoto, Silicon heterojunction solar cell with interdigitated back contacts for a photoconversion efficiency over 26%. Nat. Energy 2 (2017) 17032.

13. G. Koblmüller, S. Hertenberger, K. Vizbaras, M. Bichler, F. Bao, J-P Zhang, G. Abstreiter, Self-induced growth of vertical free-standing InAs nanowires on Si(111) by molecular beam Epitaxy, Nanotechnol. 21 (2010) 365602.

14. W. Wei, X.-Y. Bao, C. Soci, Y. Ding, Z.-L. Wang and D. Wang, Direct heteroepitaxy of vertical InAs nanowires on Si substrates for broad band photovoltaics and photodetection, Nano Lett. 9 (2009) 2926–2934.

15. J. C. Shin, P. K. Mohseni, K. J. Yu, S. Tomasulo, K. H. Montgomery, M. L. Lee, J. A. Rogers, X. Li, Heterogeneous integration of InGaAs nanowires on the rear surface of Si solar cells for efficiency enhancement, ACS Nano 6 (2012) 11074–11079.

16. A. Goodrich, P. Hacke, Q. Wang, B. Sopori, R. Margolis, T. L. James, M. Woodhouse, A wafer-based monocrystalline silicon photovoltaics road map: Utilizing known technology improvement opportunities for further reductions in manufacturing costs, Sol. Energy Mater. Sol. Cells 114 (2013) 110-135.

17. K. Masuko, M. Shigematsu, T. Hashiguchi, D. Fujishima, M. Kai, N. Yoshimura, T. Yamaguchi, Y. Ichihashi, T. Mishima, N. Matsubara, T. Yamanishi, T. Takahama, M. Taguchi, E. Maruyama, Shingo Okamoto, Achievement of more than 25% conversion efficiency with crystalline silicon heterojunction solar cell, IEEE J. Photovolt. 4 (2014) 1433-1435.

18. R. R. King, D. C. Law, K. M. Edmondson, C. M. Fetzer, G. S. Kinsey, H. Yoon, R. A. Sherif, N. H. Karam, 40% efficient metamorphic GaInP∕GaInAs∕Ge multijunction solar cells, Appl. Phys. Lett. 90 (2007) 183516.

19. J. S. Ward, T. Remo, K. Horowitz, M. Woodhouse, B. Sopori, K. Van Sant, P. Basore, Techno-economic analysis of three different substrate removal and reuse strategies for III-V solar cells, Prog. Photovolt. Res. Appl. 24 (2016) 1284–1292.

20. M. Woodhouse, A. Goodrich, H. Lee., G. P. Smestad, in: 11th Intl. Conf. Conc. Photovolt. Syst., Aix-les-Bains, France (2015).

21. R. Cariou, J. Benick, P. Beutel, N. Razek, C. Flötgen, M. Hermle, D. Lackner, S. W. Glunz, S. Member, A. W. Bett, M. Wimplinger, F. Dimroth, Monolithic two-terminal III-V//Si triple-junction solar cells with 30.2% efficiency under 1-sun AM1.5g, IEEE J. Photovolt. 7 (2017) 367–373.

22. H. Cotal, C. Fetzer, J. Boisvert, G. Kinsey, R. King, P. Hebert, H. Yoona, N. Karama, III–V multijunction solar cells for concentrating photovoltaics, Energy Environ. Sci. 2 (2009) 174–192.

23. M. A. Green, A. Ho-Baillie, H. J. Snaith, The emergence of perovskite solar cells, Nat. Photon. 8 (2014) 506–514.

24. B. O'Regan, M. Grätzel, A low-cost, high-efficiency solar cell based on dye-sensitized colloidal $TiO_2$ films, Nat. 353 (1991) 737–740.

25. A. Kojima, K. Teshima, T. Miyasaka, Y. Shirai, Novel photoelectrochemical cell with mesoscopic electrodes sensitized by lead-halide compounds, Meet. Abstr. MA2006-02 (2006) 397–397.

26. H.-S. Kim C.-R. Lee, J.-H. Im, K.-B. Lee, T. Moehl, A. Marchioro, S.-J. Moon, R. Humphry-Baker, J.-H Yum, J. E. Moser, M. Grätzel, N. G. Park, Lead iodide perovskite sensitized all-solid-state submicron thin film mesoscopic solar cell with efficiency exceeding 9%, Sci. Rep. 2 (2012) 591.

27. W. S. Yang, J. H. Noh, N. J. Jeon, Y. C. Kim, S. Ryu, J. Seo, S. I. Seok, High-performance photovoltaic perovskite layers fabricated through intramolecular exchange, Sci. 348 (2015) 1234–1237.

28. N. N. Lal, Y. Dkhissi, W. Li, Q. Hou, Y.-B. Cheng, U. Bach, Perovskite Tandem Solar Cells, Adv. Energy Mater. 7 (2017) 1602761.

29. W. Shockley, H. J. Queisser, Detailed balance limit of efficiency of p-n junction solar cells, J. Appl .Phys. 32 (1961) 510–519.

30. M. H. Futscher, B. Ehrler, Efficiency Limit of Perovskite/Si Tandem Solar Cells, ACS Energy Lett. 1 (2016) 863–868.

31. K. A. Bush, A. F. Palmstrom, Z. J. Yu, M. Boccard, R. Cheacharoen, J. P. Mailoa, D. P. McMeekin, R. L. Z. Hoye, C. D. Bailie, T. Leijtens, I. M. Peters, M. C. Minichetti, N. Rolston, R. Prasanna, S. Sofia, D. Harwood, W. Ma, F. Moghadam, H. J. Snaith, T. Buonassisi, Z. C. Holman, S. F. Bent, M. D. McGehee, 23.6%-efficient monolithic perovskite/silicon tandem solar cells with improved stability, Nat. Energy 2 (2017) 17009.

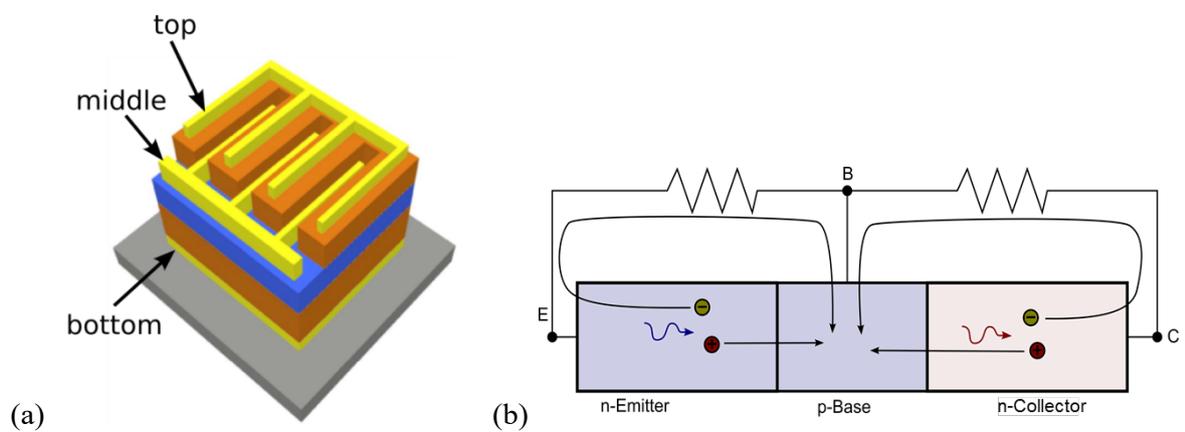

(a) (b)

**Fig. 1 (a)** Sketch of a generic 3T-HBTSC structure, where the three main n-p-n (or p-n-p) semiconductor layers are shown (orange and blue) together with their corresponding metal contacts. **(b)** Simplified layer structure and electrical representation of an n-p-n 3T-HBTSC.

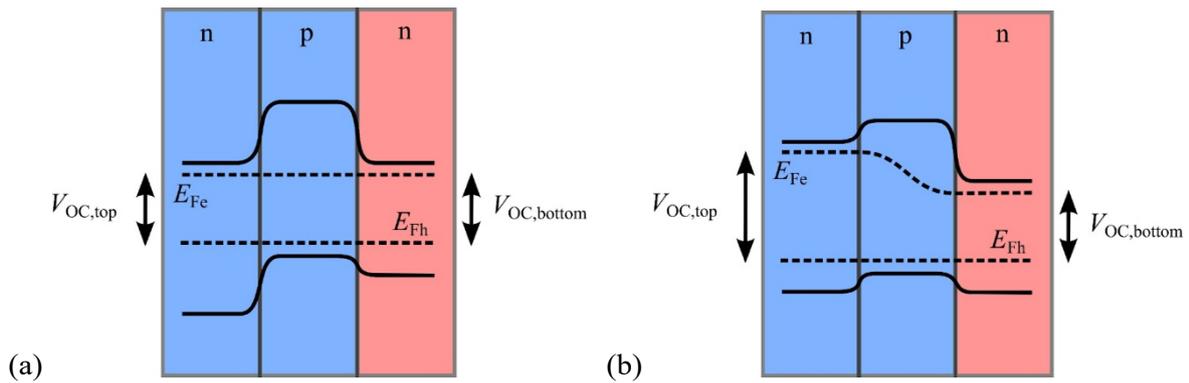

**Fig. 2** Energy band diagram of an n-p-n solar cell. **(a)** The QFL split of the p-n bottom junction extends to the top n-p junction, thus limiting the $V_{OC}$ of the top cell. **(b)** Proper 3T-HBTSC semiconductor features (e.g., bandgap, doping level and thickness) allow electron QFL bending, enabling a top subcell $V_{OC}$ that is not limited by the bottom subcells. This second behavior is achieved when the so-called injection efficiency of the equivalent n-p-n transistor approaches zero (in contrast to conventional BJTs, in which the injection efficiency must approach unity).

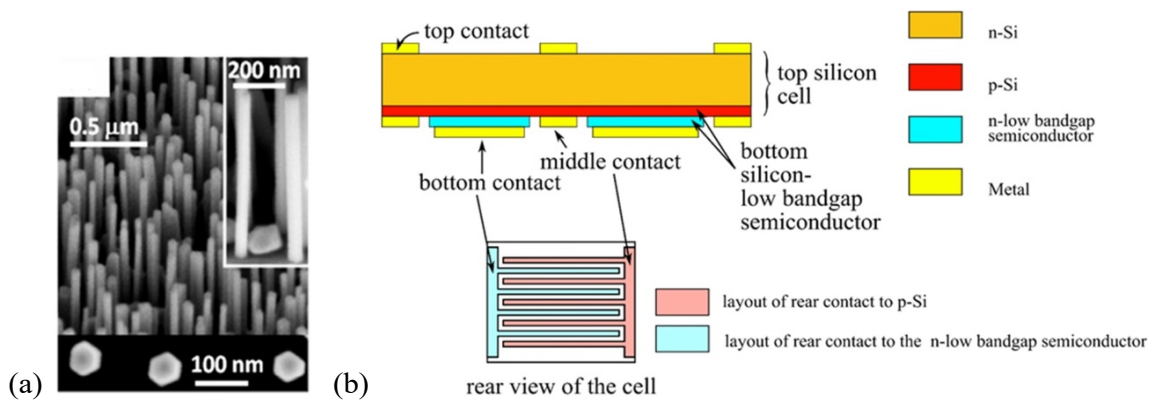

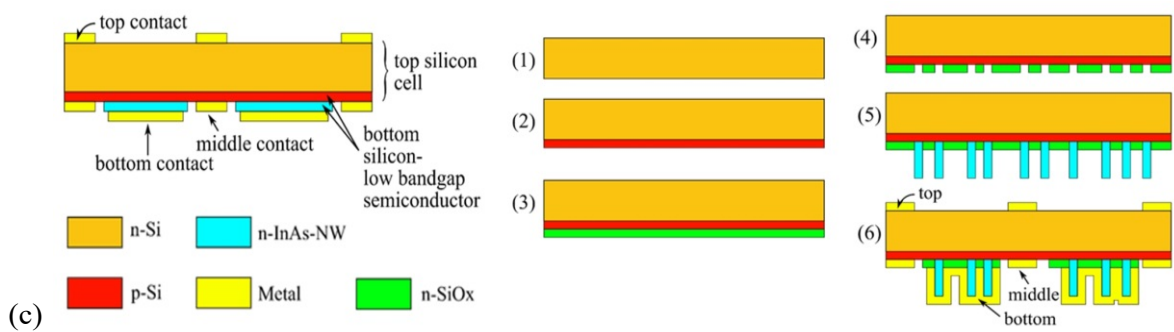

**Fig. 3 (a)** SEM image showing self-assembled NWs grown on silicon (following [13]). **(b)** Simplified layout describing the 3T-HBTSC concept implemented with a crystalline Si top junction and an In(Ga)As NW heterojunction under the Si cell. **(c)** Illustration of the main manufacturing steps of the Si/In(Ga)As NW 3T-HBTSC. Reproduced with permission [13]. Copyright 2010, IOPscience.

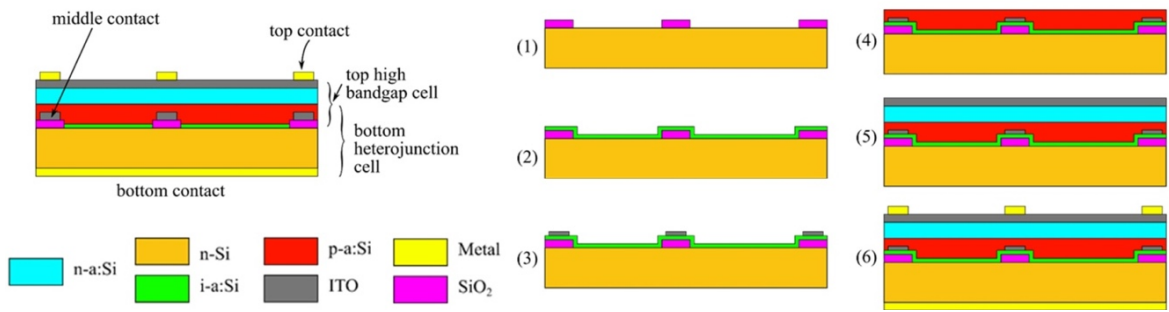

**Fig. 4** Sketch of the layer architecture of an a-Si:H/HIT 3T-HBTSC, including a step-by-step representation of the fabrication processes involved.

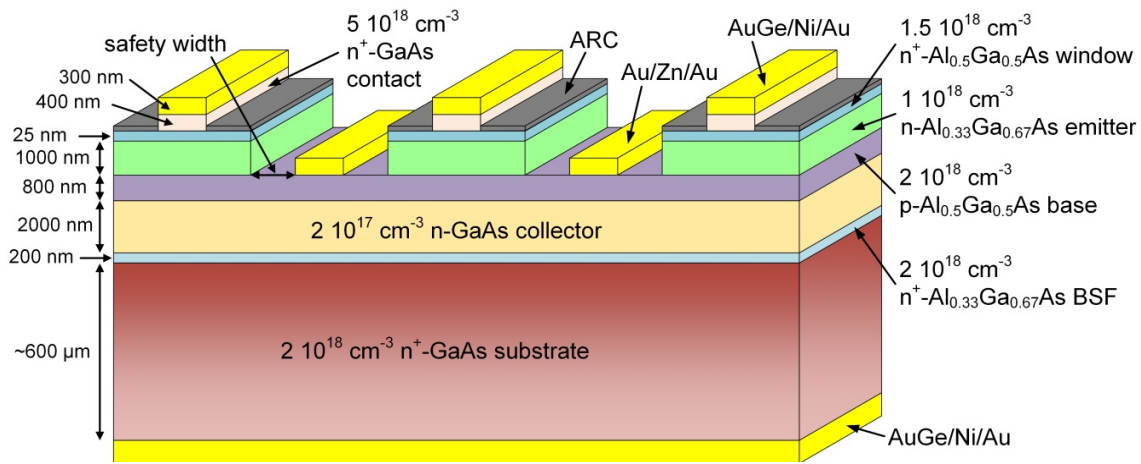

**Fig. 5** Sketch of the layer structure of an AlGaAs/GaAs 3T-HBTSC, where the main parts of the cell are indicated along with the III-V alloy stoichiometry, layer thickness and doping level.

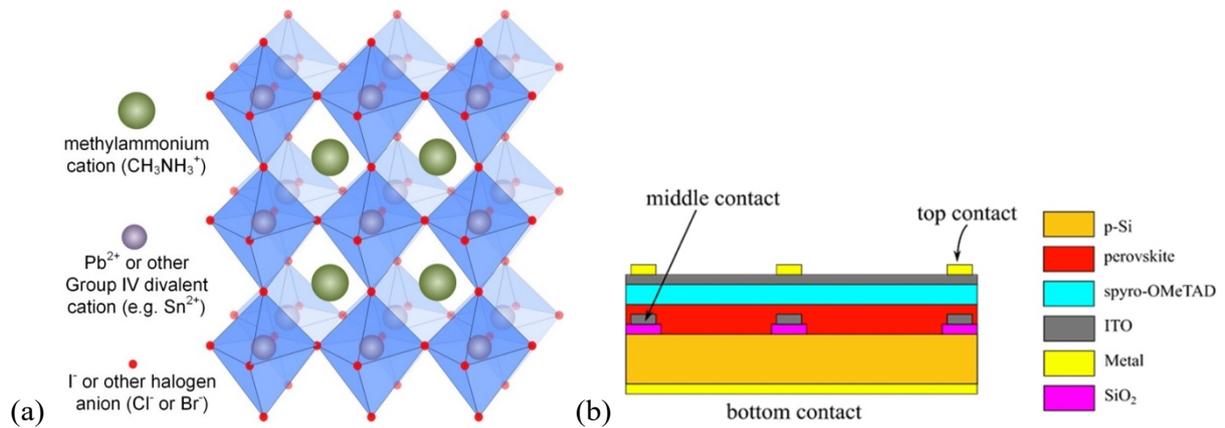

**Fig. 6 (a)** The perovskite crystalline structure most often used in PV solar cells. The cubic unit cell is composed of a methylammonium cation ($CH_6N^+$), a smaller Pb or other Group IV element cation and a halogen anion (I, Cl or Br) that bonds to both other parts. **(b)** Sketch of a 3T-HBTSC by the hybridization of a perovskite solar cell with an active Si wafer, forming a perovskite/c-Si heterojunction.